\documentstyle[amsmath,amsfonts,aps]{revtex}

\title{
New derivation of a third post-Newtonian equation 
of motion for relativistic 
compact binaries without ambiguity 
}

\author{Yousuke Itoh$\mbox{}^1$\footnote{yousuke@aei.mpg.de} 
and  
Toshifumi Futamase$\mbox{}^2$\footnote{tof@astr.tohoku.ac.jp}}
\address{
$\mbox{}^1$
Max-Planck-Institut f\"ur Gravitationsphysik, Albert-Einstein-Institut \\ 
Am M\"uhlenberg 1, Golm 14476, Germany \\ 
$\mbox{}^2$
Astronomical Institute, Graduate School of Science, Tohoku University \\
Sendai, 980-8578, Japan
}

\begin{document}
\maketitle

\begin{abstract}
A third post-Newtonian (3 PN) 
equation of motion for an inspiralling
binary consisting of two spherical compact stars with 
strong internal gravity is derived under  
harmonic coordinate condition 
using the strong field point particle limit. 
The equation of motion is complete in a sense that 
it is Lorentz invariant in the post-Newtonian perturbative sense, 
admits conserved energy of the orbital motion,  and is 
unambiguous, that is, with no undetermined coefficient. 
In this paper, we show 
explicit expressions of 
the 3 PN equation of motion and 
an energy of the binary orbital motion 
in case of the circular orbit 
(neglecting the 2.5 PN radiation reaction effect) 
and in the center of the mass frame.
It is argued that 
the 3 PN equation of motion we obtained is physically 
unambiguous. Full details will be reported elsewhere.
\end{abstract}

\begin{flushleft}
PACS Number(s): 04.25.Nx,04.25.-g 
\end{flushleft}

\newcommand{\pa}{\partial}

Renewed attention has been paid to a high order post-Newtonian 
equation of motion governing  inspiralling compact binaries 
in the context of the efforts for direct detection of 
gravitational waves \cite{CutlerEtAL93,Blanchet02b}.
It is well-known that detectability of the gravitational waves  
emitted by the 
binaries and 
quality of measurements of astrophysical information (e.g. masses) 
depend on accuracy of theoretical knowledge 
of the waveforms \cite{CutlerEtAL93}, and hence partly of   
dynamics of the binaries.

The 
3 PN approximation has been a subject 
of much discussion because of its ambiguity reported originally 
in Jaranowski and Sch\"afer \cite{JS3PNHamiltonian}. 
In fact, the 3 PN ADM Hamiltonian in the ADM-type gauge obtained 
in \cite{JS3PNHamiltonian} has two undetermined coefficients  
($\omega_{{\rm kinetic}}$ and $\omega_{{\rm static}}$) 
and the 3 PN equation of motion in the harmonic gauge 
derived by Blanchet and Faye \cite{BF3PNEOM} has one 
coefficient $\lambda$ undetermined within their framework. 
Both groups have used Dirac delta distributions, which 
cause divergences in general relativity, to express 
the point particles and inevitably they have resorted to mathematical 
regularizations. Damour {\it et al.} \cite{DJS01a} pointed out that 
the undetermined coefficients may arise due to unsatisfactory 
 features of the regularizations they have used in 
\cite{JS3PNHamiltonian,BF3PNEOM}. Indeed, 
using the dimensional regularization,  
the work \cite{DJS01a} have succeeded in determining both of 
the coefficients, namely,   
$\omega_{{\rm static}} = 0$, which means $\lambda=-1987/3080$ 
via relationship established in \cite{DJSABF01}. 
($\omega_{{\rm kinetic}}$ is related with Lorentz invariance 
and was fixed in \cite{DJS01a,DJS00}. 
Blanchet and Faye have developed a Lorentz invariant 
Hadamard Patie Finie regularization \cite{BF00b,BF01b} and 
do not have 
any ambiguity 
other than $\lambda$.)

In gravitational wave data analysis, 
the reduction of 
predictability of the equation of motion due to
the  undetermined coefficient can become a problem. 
In fact,  
the 3.5 PN phase evolution equation and 
luminosity \cite{3hWaveTemplate} unfortunately 
have four undetermined coefficients, one of which 
is $\lambda$.

Theoretically, a use of Dirac delta distributions and 
inevitable regularization should be verified in 
some manner. 
The perfect (physical) agreements among the results 
obtained by various authors with various methods 
\cite{2hPNApproximation,IFA01,PW02} 
give a direct theoretical confirmation 
of the 2.5 PN result first derived by 
Damour and Deruelle \cite{DD81}. 
It is important to achieve 3 PN 
iteration without introducing singular sources 
to derive unambiguous result and 
support the previous 3 PN works which have used 
Dirac delta distributions.

Based on our previous papers \cite{IFA01,IFA00}, we derive 
a 3 PN equation of motion 
for two spherical compact stars 
in harmonic gauge 
without introducing 
singular sources. 
Instead, 
we apply the strong field point particle limit \cite{Futamase87} 
to deal with strong internal
gravity of the stars. 
Our derivation 
is satisfactory in a sense that  
the equation admits conserved energy, is Lorentz invariant, and 
is unambiguous.
In this paper, we shall show both of 
the 3 PN equation 
of motion and an associated 3 PN energy of the orbital motion 
in the center of mass frame and in the case of circular orbit.

Below, we shall explain briefly our yet another derivation 
of a 3 PN equation of motion. Since this method is different from 
others, we mention some details specific to 
our method at the 3 PN order. 
After deriving an invariant 
energy of the binary orbital motion, we shall compare it with that 
derived by Blanchet and Faye and fix the $\lambda$ parameter.
Full explanation of our method including  
computational details will be reported in \cite{Itoh04}. See also 
\cite{IFA01,IFA00}.

We write explicitly the post-Newtonian expansion 
parameter, $\epsilon$, which represents the smallness of the orbital 
velocities of the stars. The mass scales as $\epsilon^2$ from 
the post-Newtonian scaling. Then the 
{\it strong field point particle limit} 
\cite{Futamase87} is achieved by setting  
the radius of the star to scale at the same rate 
as its mass, $\epsilon^2$.
The scalings of the mass and the radius enable us to incorporate 
in the post-Newtonian limit ($\epsilon \rightarrow 0$) 
a limit of a regular point particle whose internal gravity, 
$\sim$ the mass over the radius, is strong irrelevantly to 
$\epsilon$. 

We derive an equation of motion via surface integrals of 
the gravitational energy momentum flux going through a sphere surrounding 
the star. For this method, we prepare two spheres 
$B_A(\tau) \equiv \{x^k| |\vec x - \vec z_A(\tau)|\le \epsilon R_A\}$  
($A$ labels the two stars) 
on the $\tau=$ constant surface, where  
$\tau$ is the time coordinate in the near zone. 
$B_A$, called the body zone, 
is centered at the star A's representative point $z_A^i(\tau)$ and 
has a radius $\epsilon R_A$ where $R_A$ is an arbitrary constant 
and smaller than the orbital separation but larger than the radii  
of the stars. We make 
the body zone radius shrink proportionally to $\epsilon$ 
in the near zone coordinate $(\tau,x^i)$ 
to ensure that the field on the body zone boundary 
due to the star is obtained by multipole expansion when the 
$\epsilon$ zero limit is taken.

The $l$-th multipole moments of the star $A$ 
denoted by $I_A^L(\tau)$, including its mass,  
are defined as volume integrals over $B_A$ 
of $\Lambda^{\tau\tau} \equiv 
-g(T^{\tau\tau} + t_{LL}^{\tau\tau})  
+ \chi^{\tau\tau\alpha\beta}\mbox{}_{,\alpha\beta}$ 
where $g, T^{\mu\nu}$ and $t_{LL}^{\mu\nu}$ 
are the determinant of the metric $g_{\mu\nu}$, 
the matter stress energy tensor and  
the Landau-Lifshitz pseudo tensor.   
$\chi^{\mu\nu\alpha\beta}\mbox{}_{,\alpha\beta}$ arises since 
we use the wave operator of the flat spacetime instead of 
that of the curved spacetime when we solve the harmonically 
relaxed Einstein equation. 
\begin{eqnarray}
I_A^L \equiv \epsilon^2 \int_{B_A(\tau)}d^3\alpha_A 
\Lambda^{\tau\tau}\alpha_A^L, 
\end{eqnarray}
where we introduced the body zone coordinate 
$\alpha_A^i = \epsilon^{-2}(x^i - z^i_A(\tau))$ and   
multi-indices notation $L=i_1\cdots i_l$ 
($l \ge 0:$ integer). $\alpha^L_A = \alpha_A^{i_1}\cdots
\alpha_A^{i_l}$.
The mass $m_A \equiv \lim_{\epsilon \rightarrow 0} P^{\tau}_A$ with  
$ P^{\tau}_A \equiv I^0_A$ 
so defined would be 
the ADM mass if the companion star were absent and 
the body zone radius is taken to be infinite 
in the body zone coordinate.

In deriving a 3 PN equation of motion, it is important 
in our formalism to notice 
that the body zone $B_A$ is a sphere in the frame where 
the star $A$ orbits, but $B_A$ is {\it not} a sphere 
in the generalized Fermi frame \cite{AB86} where 
the star $A$ is at rest and   
the effect of the gravitational field due to the 
companion star 
is removed (as much as possible) except for, namely, the tidal effect.
We define stars to be spherical in the generalized Fermi frame. 
Now, we define the ``intrinsic'' multipole moments 
$\hat I_A^L(\hat \tau)$ on   
$\hat \tau=$ constant surface 
in the generalized Fermi coordinate $(\hat \tau, \hat x^i)$ 
as a volume integral over a sphere $\hat B_A(\hat \tau)$ centered 
at the same world event and 
with the same radius $\epsilon R_A$ as $B_A(\tau)$.  
For example, the difference between the symmetric tracefree 
quadrupole moments is defined as 
$
\epsilon^{8}
\delta I_A^{<ij>} \equiv \epsilon^8I_A^{<ij>}-\epsilon^8\hat I_A^{<ij>}=
\int_{B_A(\tau)}d^3 y_A 
\Lambda^{\tau\tau} y_A^{<i} y^{j>} 
-  
\int_{\hat B_A(\hat \tau)}d^3\hat y_A 
\Lambda^{\hat \tau \hat \tau}\hat y_A^{<i}\hat y_A^{j>} 
$ where $<..>$ denotes symmetric tracefree operation on the indices 
between the brackets. For a 
``spherically symmetric'' compact stars, the ``intrinsic'' 
tracefree quadrupole moment $\hat I_A^{<ij>}$ vanishes. 
Up to the 3 PN iteration of the gravitational field, 
$\delta I_A^{<ij>}$ arise mainly due to the Lorentz contraction 
and can be evaluated as a surface integral over $\partial B_A$.   
\begin{eqnarray}
\delta I_A^{<ij>} = 
\epsilon^{-6} \frac{1}{2}v_A^kv_A^l
\oint_{\partial B_A}dS_ky_A^{l}y_A^{<i}y_A^{j>}\Lambda^{\tau\tau}
+O(\epsilon^4)
= 
- \epsilon^2 \frac{4}{5}m_A^3v_A^{<i}v_A^{j>}
+O(\epsilon^4).  
\label{eq:deltaIij}
\end{eqnarray} 
In our formalism, it is possible to derive the 3 PN field 
for an isolated star (by taking limit where the mass 
of the companion star is zero). $\delta I_A^{<ij>}$ is a  
necessary term to obtain the correct expression of the 
3 PN field for such a system.
Although other multipole moments defined over $B_A$ possibly hide 
purely monopole terms, only the quadrupole moment is found 
to be relevant up to the 3 PN order.
Clean extraction of monopole terms from the 
multipole moments defined in our previous works is a problem 
at the 3 PN order specific to our formalism. 
Blanchet and Faye on the other hand elaborate their 
generalized Hadamard Partie Finie regularization \cite{BF00b} 
in a Lorentz 
invariant manner \cite{BF01b}, 
and have properly taken into account of special relativistic 
kinematic effects including 
the Lorentz contraction.

Now, let us briefly explain our derivation of the equation of motion.
The local conservation law of the total energy gives 
an evolution equation for  four momentum of the 
star and 
relationships among the multipole moments, namely, 
momentum-velocity relation. The last read as    
$P_A^i = P_A^{\tau}v_A^i + Q_A^i + \epsilon^2 d D_A^i/d \tau$ 
where $P_A^i \equiv \epsilon^2\int_{B_A}d^3\alpha_A\Lambda^{\tau i}$ 
and $P_A^{\tau}$ are the three momentum and 
the energy of the star $A$. 
$Q_A^i \equiv \epsilon^{-4}
\oint_{B_A}dS_k(\Lambda^{\tau k}-v_A^k\Lambda^{\tau\tau})y_A^i$ 
arises since 
the (pseudo-)stress energy momentum of the field extends outside of the 
star \cite{IFA01}.    
$Q_A^i$ can be evaluated explicitly and do contribute the 
3 PN velocity momentum relation. 
We can define the representative point 
$z_A^i(\tau)$  
of the star $A$ 
by specifying the dipole moment of the star, $D_A^i \equiv I_A^i$ (, e.g., 
$z_A^i(\tau)$ corresponding to a condition $D_A^i = 0$ 
may be called the 
center of mass of the star $A$ from an analogy of the Newtonian 
dynamics). The relationship between the energy $P_A^{\tau}$ 
and the mass $m_A$ can be obtained by integrating functionally 
the evolution equation of $P_A^{\tau}$, which is expressed as 
surface integrals and can be evaluated explicitly up to the 3 PN order,
 in the form as $P_A^{\tau} = m_A[1 + O(\epsilon^2)]$.

Combining  
the mass energy relation, the momentum velocity relation, 
and the evolution equation for the four momentum,   
we obtain the general form of the equation of motion 
\cite{IFA01};
\begin{eqnarray}
m_A \frac{dv_A^i}{d\tau} &=&
 -\epsilon^{-4}
 \oint_{\pa B_A} dS_k \Lambda^{ki}
+ \epsilon^{-4}
v_A^k \oint_{\pa B_A} dS_k \Lambda^{\tau i}
\nonumber\\
&&
 +\epsilon^{-4}
 v_A^i \left( \oint_{\pa B_A} dS_k \Lambda^{k\tau}
-v_A^k \oint_{\pa B_A} dS_k \Lambda^{\tau\tau} \right)
\nonumber\\
&&-\frac{dQ_A^i}{d\tau}  - \epsilon^2 \frac{d^2 D_A^i}{d \tau^2}   
+ 
(m_A - P_A^{\tau}) \frac{dv_A^i}{d\tau}
\label{generaleom}
\end{eqnarray} 
Note that 
$\Lambda^{\mu\nu} = (-g)t_{LL}^{\mu\nu} 
+ \chi^{\mu\nu\alpha\beta}\mbox{}_{,\alpha\beta}$ on 
$\pa B_A$, since $\partial B_A$ is well outside the 
star by construction of the body zone. 
The acceleration in the right hand side of 
Eq. (\ref{generaleom}) should be understood to be  
lower order acceleration than in the left hand side.
The terms in the right hand side of Eq. (\ref{generaleom}) 
are completely expressed as surface integrals over the 
body zone boundary except for $D_A^i$ to be specified. 
The surface integral approach enables us to derive 
an equation of motion irrelevant to the internal structure 
of the star (Effects of the star's internal structure 
on the orbital motion  
such as tidally induced multipole moments  
appear through the field and hence the integrand 
$\Lambda^{\mu\nu}$). 
The scaling of the body zone radius $\epsilon R_A$ 
ensures that we have an equation of motion for 
compact stars.

The field equation coupled to the matter 
equations mentioned above is the integrated relaxed 
Einstein equation under harmonic gauge 
($h^{\mu\nu} \equiv \eta^{\mu\nu} - \sqrt{-g}g^{\mu\nu}$ where 
$\eta^{\mu\nu} = 
{\rm diag}(-\epsilon^2,1,1,1)$. The harmonic condition is 
then $h^{\mu\nu}\mbox{}_{,\nu} = 0$) 
\begin{equation}
h^{\mu\nu}(\tau,x^i)=4 \int_{C(\tau,x^i)} d^3y 
\frac{\Lambda^{\mu\nu}(\tau-\epsilon|\vec x-\vec y|, y^k; \epsilon)} 
{|\vec x-\vec y|},
\label{IntegratedREE}
\end{equation}
We split the flat light cone $C(\tau,x^i)$ into four parts;
two body zones $B_A$, near zone outside the body zones $N/B$ 
surrounding the binary and the 
far zone outside the near zone. 
For $B_A$ and $N/B$ contributions to the field $h^{\mu\nu}(\tau,x^i)$, 
we expand the retarded field about the near zone time $\tau$. 
Then multipole expansion of the star is used to 
evaluate the two body zone contributions. 
The $N/B$ contribution is basically evaluated 
with help of super-potentials (a super-potential 
here means a particular solution valid in $N/B$ 
of a Poisson equation with a non-compact support source 
in $N/B$). 
Unfortunately, it was not possible to find all the necessary 
super-potentials explicitly at the 3 PN order. 
For integrands in 
Eq. (\ref{IntegratedREE}) such that we could not find  
super-potentials 
in closed forms, 
after making the retarded expansion 
we leave the Poisson integrals unevaluated and 
substitute the (not-integrated) 
field into $\Lambda^{\mu\nu}$  
in Eq. (\ref{generaleom}). Then  
we perform the surface integrals 
in Eq. (\ref{generaleom}) 
(with respect to the spatial variable $x^i$ in
Eq. (\ref{IntegratedREE})) 
first and 
next perform the remaining volume integrals  
(with respect to the spatial variable $y^i$ in 
Eq. (\ref{IntegratedREE})). 
In other words, we extract the parts of the field necessary 
to derive the equation of motion by interchanging the order of the 
integrations in Eq. (\ref{generaleom}) and 
Eq. (\ref{IntegratedREE}). As a check, we applied this method 
on the integrands for which the necessary super-potentials can 
be derived 
in closed forms, 
and found that both methods give the same result.
Finally, 
we have dealt with the integration over the far zone
using the DIRE method \cite{PW02,DIRE} and found that 
it does not 
contribute to the 3 PN equation of motion \cite{PW02,BD88}.

Using the method mentioned above, we obtain a 3 PN equation 
of motion for a two spherical compact stars binary. 
We here present the 3 PN relative acceleration 
in the case of the circular orbit 
and in the center of mass frame, which is an appropriate equation 
to inspiralling binaries. 
\begin{eqnarray}
\frac{d V^i}{d\tau} &=& - \Omega_{\ln}^2 r_{12}^i + \epsilon^5 
\mbox{}_{{\rm 2.5 PN}}A^i,  
\label{eq:3PNeom}
\end{eqnarray}
where $V^i = v_1^i - v_2^i$ is the relative velocity 
and  
$\mbox{}_{{\rm 2.5 PN}}A^i$ is the relative 
acceleration at 
the 2.5 PN order (the radiation reaction term). 
The 3 PN orbital angular frequency $\Omega_{\ln}$ is,  
(for comparison, we adopt similar notations as 
in \cite{BF3PNEOM})
\begin{eqnarray}
m^2 \Omega^2_{\ln} &=& 
\gamma^3\left[
1 + \epsilon^2\gamma(-3+\nu)
+\epsilon^4\gamma^2 \left(6+\frac{41}{4}\nu + \nu^2\right)
\right. 
\nonumber \\
\mbox{} &&
\left. 
+\epsilon^6\gamma^3 \left(-10
+ \nu\left\{
-\frac{2375}{24}+\frac{41\pi^2}{64}+
22\ln\left(\frac{r_{12}}{R_0}\right)
\right\}
+\frac{19}{2}\nu^2 + \nu^3 
\right)
\right]
+ O(\epsilon^7),
\label{eq:3PNorbitalangularfrequency} 
\end{eqnarray}
where 
$m = m_1 + m_2$, $\nu = m_1m_2/m^2$, $\gamma = m/r_{12}$, 
and $\ln R_0 = (m_1 \ln (\epsilon R_1) + 
m_2 \ln (\epsilon R_2))/m$. 
In Eq. (\ref{eq:3PNeom}) with (\ref{eq:3PNorbitalangularfrequency}), 
the representative 
point of the star $A$, $z_A^i(\tau)$, is defined by setting 
$D_A^i = \epsilon^4 \delta_{A Q}^i = 
 - 86 \epsilon^4 m_A^3 \mbox{}_Na_A^i/9$, where 
$\mbox{}_Na_A^i$ is the Newtonian acceleration.  
This choice makes the three momentum $P^i_A$ parallel to 
the velocity $v_A^i$.
We note that there is no arbitrary parameter 
other than the body zone radii $\epsilon R_A$ in the 
3 PN relative acceleration. 
We here note that it is not allowed to 
fix the $\lambda$ parameter by 
comparing  
Eq. (\ref{eq:3PNorbitalangularfrequency}) 
with the corresponding result of 
Blanchet and Faye \cite{BF3PNEOM},  since  
the harmonic condition both groups have used 
does not fix the gauge completely \cite{Itoh04} and 
the expression of the 3 PN orbital angular frequency in terms of 
the coordinate distance $m/r_{12}$, 
Eq. (\ref{eq:3PNorbitalangularfrequency}), 
is gauge dependent. 
(From the same reason, we can not fix $\lambda$ using 
Eq. (\ref{eq:ElninGamma}) below.)

We can remove away $\epsilon R_A$ dependence from the 3 PN 
relative acceleration, Eq. (\ref{eq:3PNeom}),
physically by a suitable redefinition 
of the representative points of the stars. 
In fact, by setting  
\begin{eqnarray}
D_{A,{\rm New}}^i &=& \epsilon^4 \delta_{A Q}^i  
- \epsilon^4\frac{22}{3}m_A^3\mbox{}_Na_A^i
\ln\left(\frac{r_{12}}{\epsilon R_A}\right)
\label{eq:redefinitionOfZ}
\end{eqnarray} 
we obtain the 3 PN relative acceleration free from any 
arbitrary parameter 
\begin{eqnarray}
\frac{d V^i}{d\tau} &=& - \Omega^2 r_{12}^i + \epsilon^5 
\mbox{}_{{\rm 2.5 PN}}A^i,  
\label{eq:3PNEOMwithoutLOG}
\end{eqnarray}
with $m^2\Omega^2 = 
m^2\Omega^2_{\ln} - 22 \epsilon^6 \gamma^6\nu \ln(r_{12}/R_0)$.
This observation 
in fact is the case in general cases (
i.e., in general orbits {\it not} in the center of mass frame);   
The 3 PN equation of motion in general cases  
we have derived is physically 
free from any ambiguity.

A reason why we are concerned with $\ln \epsilon R_A$ dependence 
is the following. 
Blanchet and Faye have introduced four 
arbitrary parameters in their regularization procedure, two 
of which appear in the regularization of the field having 
two singular points, and the others appear in the regularization 
of equations of motion for those two points. They showed that 
the two of those parameters can be gauged away, while the other 
two were consumed to make their equations of motion conservative 
(modulo the radiation reaction effect), 
and they found there remained one and only one parameter, $\lambda$, 
although relationship between energy conservation and regularization 
parameters associated with point particle description is not clear. 
Our redefinition of the representative points 
(\ref{eq:redefinitionOfZ}) corresponds to their gauge transformation.
Then, their observation makes us check if 
it is physically allowed to remove the 
$\ln \epsilon R_A$ dependence in our 3 PN equation of motion, 
since we introduced only two 
arbitrary parameters $\epsilon R_A$ and we have no freedom to make our 
equation motion conservative by adjusting these two parameters if 
we remove them away. Thus, we have two problems to be solved 
in our method; removal of $\ln \epsilon R_A$ and an energy conservation. 
For lack of space, here we show some facts 
which support naturality of Eq. (\ref{eq:redefinitionOfZ}). 
The energy conservation problem will be addressed 
in \cite{Itoh04}, there we shall show our equation of motion 
and an associated conserved energy of the binary orbital motion 
in general cases.

Let us consider the harmonic 
condition.   
\begin{eqnarray}
0= h^{\tau \mu}\mbox{}_{,\mu} &=& 
4 \epsilon^4 
\sum_{A=1,2}
\left[
%\frac{\dot P_A^{\tau}}{r_A} 
\frac{1}{r_A}\frac{d P^{\tau}_A}{d\tau} 
+ \frac{r_A^i}{r_A^3}\left(P_A^{\tau}v_A^i 
%+ \epsilon^2 \dot D_A^i 
+ \epsilon^2 \frac{d D_A^i}{d \tau} 
- P_A^i \right) 
\right.
\nonumber \\
\mbox{} &&+ \left. \sum_{A=1,2}
\oint_{\pa B_A}\frac{dS_i}{|\vec x - \vec y|}
\left(\Lambda^{\tau i} - v_A^i\Lambda^{\tau\tau}\right)
+ \cdots
\right], 
\\
0= h^{i \mu}\mbox{}_{,\mu} &=&
\left[
\epsilon^4 
\sum_{A=1,2} 
%\frac{\dot P_A^i}{r_A} 
\frac{1}{r_A}\frac{d P_A^i}{d\tau} 
+ \sum_{A=1,2}
\oint_{\pa B_A}\frac{dS_j}{|\vec x - \vec y|}
\left(\Lambda^{ij} - v_A^j\Lambda^{\tau i}\right)
+ \cdots
\right] ,
\label{eq5-69} 
\end{eqnarray}
where ``$\cdots$'' are irrelevant terms. These equations 
are manifestation of the fact that the harmonic condition 
is consistent with the evolution equation of $P_A^{\tau}$, 
the momentum-velocity relation, and the equation of motion 
(and relations among higher multipole moments, hidden in 
``$\cdots$'').  Thus, if logarithmic 
dependence of $\epsilon R_A$ arises from the 
equation of motion (essentially the 
second term of Eq. (\ref{eq5-69})),  
$P_A^i$ must have the same 
logarithmic dependence (times minus sign) to ensure 
harmonicity. This and the momentum velocity relation 
in turn mean $P_A^{\tau}$, 
$v_A^i = d z_A^i/d\tau $ or $D_A^i$ 
have corresponding logarithmic dependence.  
We found that $P_A^{\tau}$ have no logarithm  
up to the 3 PN order. 
Therefore $z_A^i$ or $D_A^i$ should have 
logarithms. This is consistent with the fact 
that a choice of $D_A^i$ determines $z_A^i$. 
$z_A^i$ depends on logarithms if the old choice 
is taken, while it does not if our new choice is 
taken.

The second fact which supports our interpretation is 
as follows. 
We find that  the near zone dipole moment $D_N^i$ 
defined by a volume integral of $\Lambda^{\tau\tau}y^i$  
becomes 
\begin{eqnarray}
\epsilon^2 D_N^{i} \equiv \epsilon^{-4}
\int_{N}d^3y \Lambda^{\tau\tau}y^i = 
\sum_{A=1,2}P_A^{\tau} z_A^i + \epsilon^2
\sum_{A=1,2}D_A^{i} + \epsilon^{-4} \int_{N/B}d^3y \Lambda^{\tau\tau}y^i. 
\end{eqnarray}
Then if we take the old choice of $D_A^i$, 
the volume integral becomes   
\begin{eqnarray}
&&
\int_{N/B}d^3y \Lambda^{\tau\tau}y^i =  
\epsilon^4\frac{22}{3}\sum_{A=1,2}
m_A^3 \mbox{}_Na_A^i \ln \left(\frac{r_{12}}{\epsilon R_A}\right) 
+ \cdots,  
\end{eqnarray}
where terms denoted by ``$\cdots$'' are independent of $R_A$. 
Notice that the near zone dipole moment can be freely 
determined, say, $D_N^i =0$, since we can define the origin 
of near zone freely in general \cite{nearzonedipole}. 
By taking temporal derivatives 
of $D_N^i$ twice, we see that $D_{A{\rm New}}^i$ gives the 
definition of $z_A^i(\tau)$ in terms of which 
 the 3 PN equation of motion is independent of 
$\epsilon R_A$.

Finally, we show the 3 PN conserved energy  
(neglecting the 
2.5 PN radiation reaction force) of the circular 
orbital motion in the center of mass frame. 
Using Eq. (\ref{eq:3PNeom}), we have 
\begin{eqnarray}
E_{\ln}(\gamma) &=& - \frac{m \nu\gamma}{2}
\left[
1 
+ 
\epsilon^2 \gamma \left(- \frac{7}{4} + \frac{\nu}{4}\right)
+
\epsilon^4 \gamma^2 \left(
-\frac{7}{8} + \frac{49}{8}\nu + \frac{1}{8}\nu^2 
\right)
\right.
\nonumber \\
\mbox{} &&+ \left. 
\epsilon^6\gamma^3
\left(
-\frac{235}{64} + \frac{27}{32}\nu^2 + \frac{5}{64}\nu^3 
+ \nu\left\{
\frac{10141}{576}
-\frac{123\pi^2}{64}
+\frac{22}{3}\ln\left(\frac{r_{12}}{R_0}\right)
\right\}
\right)
\right] 
+O(\epsilon^7)
\label{eq:ElninGamma}
\end{eqnarray} 
In terms of $x = (m\Omega_{\ln})^{2/3}$ 
we obtain the 3 PN energy in an invariant form 
\begin{eqnarray}
E_{\ln}(x) &=& 
- \frac{m \nu x}{2}
\left[
1 + \epsilon^2\left(-\frac{3}{4}-\frac{1}{12}\nu\right)x 
+ 
\epsilon^4
\left(-\frac{27}{8}+\frac{19}{8}\nu - \frac{1}{24}\nu^2\right)
x^2 
\right.
\nonumber \\
\mbox{} &&+ \left. 
\epsilon^6
\left(
- \frac{675}{64} + 
\left\{
\frac{34445}{576}-\frac{205\pi^2}{96}
\right\}\nu
-\frac{155}{96}\nu^2
-\frac{35}{5184}\nu^3
\right)
x^3
\right] + O(\epsilon^7). 
\label{finalresult}
\end{eqnarray}
Similarly, using Eq. (\ref{eq:3PNEOMwithoutLOG}), 
we have $E(\gamma) = E_{\ln}(\gamma) 
+ (11/3)\epsilon^6m\nu^2\gamma^4\ln(r_{12}/R_0)$. 
Here we note that the difference between $E(\gamma)$ and  
$E_{\ln}(\gamma)$ is merely due to the redefinition of the 
dipole moments (or equivalently, a coordinate transformation 
under the harmonic coordinate condition).
The invariant energy $E(x)$ is the same as $E_{\ln}(x)$ 
but with $x=(m\Omega_{\ln})^{2/3}$ replaced with 
$x=(m\Omega)^{2/3}$. This third fact that 
the energy has the same form for both 
definitions of the representative points of the 
stars when we write the energy in terms of the 
orbital angular frequency which is an observable 
supports that the apparent body zone radii dependence 
of the 3 PN relative acceleration has no physical  
effect on the orbital motion. 

We have thus derived a 3 PN equation of motion which 
takes account of strong internal gravity and avoids 
any ambiguity.
Comparing our result, Eq. (\ref{finalresult}), 
with the corresponding result in \cite{BF3PNEOM}, 
we determine 
the coefficient undetermined in the Blanchet and 
Faye 3 PN equation of motion as 
$\lambda = -1987/3080$. This value of $\lambda$ 
is consistent with the result of Damour, Jaranowski, 
and Sch\"afer \cite{DJS01a}. Thus, our result 
(indirectly) 
validates 
their use of the dimensional regularization in the 
ADM Hamiltonian approach in the ADMTT gauge. 
Finally, we note that Blanchet {\it et al.} 
\cite{BDEF03} have recently obtained the same 
value of $\lambda$, who computed a 3 PN equation 
of motion in the harmonic gauge 
using dimensional regularization.

\section*{Acknowledgments}
Y. I. was partly supported 
by JSPS Research Fellowships for Young Scientists.
The authors would like to acknowledge H. Asada for fruitful 
discussion and comments. 
Extensive use has been made of the softwares 
Mathematica and MathTensor.


\begin{thebibliography}{999}
\bibitem{nearzonedipole}
We here do not discuss an appropriate definition of 
the center of mass frame of the binary system.
\end{thebibliography}
\end{document}